\documentclass[twocolumn,secnumarabicaps, prc]{revtex4}
\usepackage[dvips]{graphicx}
\usepackage{amsmath,amssymb}
\usepackage{bm}
\makeatletter

\@addtoreset{equation}{section}
\makeatother
\begin{document}
\title{Some forgotten features of the Bose-Einstein Correlations\footnote{Talk delivered
by G.Wilk at the International Workshop {\it Relativistic Nuclear
Physics: from Nuclotron to LHC energies}, Kiev, June 18-22, 2007,
Ukraine.} }
\author{G.A.Kozlov}
\affiliation{Bogoliubov Laboratory of Theoretical Physics, JINR,
Dubna, Russia}
\author{O.Utyuzh and G. Wilk}
\email{wilk@fuw.edu.pl}
\affiliation{The Andrzej So{\l}tan Institute for Nuclear Studies, Ho\.{z}a 69, 00681,
Warsaw, Poland}
\author{Z.W\l odarczyk}
\affiliation{Institute of Physics, \'Swi\c{e}tokrzyska Academy,
         Kielce, Poland}

\date{\today}

\begin{abstract}
Notwithstanding the visible maturity of the subject of
Bose-Einstein Correlations (BEC), as witnessed nowadays, we would
like to bring to ones attention two points, which apparently did
not received attention they deserve: the problem of the choice of
the form of $C_2(Q)$ correlation function when effects of partial
coherence of the hadronizing source are to be included and the
feasibility to model effects of Bose-Einstein statistics, in
particular the BEC, by direct numerical simulations.
\end{abstract}

\maketitle

\section{Introduction}

The subject of Bose-Einstein Correlations (BEC) is so matured
nowadays that (almost) everything seems to be already answered
and/or understood and what remains is just to systematically
deduce from the experimental data information on the
spatio-temporal structure of the hadronizing source. That is the
main reason of interest in BEC.  Nevertheless, we would like to
bring attention to two points which, in our opinion, are still
worth of debate or, at least, worth to be remembered. They are:
\begin{itemize}
\item What is the {\it proper form} of the two-particle
correlation function $C_2(Q)$ in the case when one wants to
account for the effects of the possible partial coherence of the
hadronizing source \cite{KUW}? \item Is it possible to {\it model}
numerically effects of Bose-Einstein statistics (BE), in
particular BEC, and in what way \cite{UWW}?
\end{itemize}
Since we address mainly readers already acquainted with the
subject of BEC, no special introduction is offered. All necessary
material can be found in \cite{KUW,UWW} and references therein.
The above mentioned two points will be addressed in two Sections
that follow. We close with short summary.

\section{Which form of correlation function is the correct one?}

In most cases, when analyzing experimental data on BEC and
discussing phenomenological models, one uses the following form
for the two body correlation function ($Q = \sqrt{-\left(p_1 -
p_2\right)^2}$ and $p_{1,2}$ are four-momenta of the observed
identical bosons):
\begin{equation}
C_2(Q)= 1 + \lambda \cdot \Omega(Q\cdot r). \label{eq:first}
\end{equation}
Here  $r_{\mu}$ is a $4$-vector parameter,  such that
$\sqrt{(r_{\mu})^2}$ has dimension of length making the product
$Q\cdot r = Q_{\mu} r_{\mu}= q$ dimensionless. It is usually
regarded as representing dimension of the hadronizing source, but
in reality it represents the mean distance between the emission
points of the two particles considered \cite{Zajc}. Parameter
$\lambda \in (0,1)$, from the theoretical point of view, is
usually understood as the degree of chaoticity of the source:
totally chaotic source has $\lambda =1$ and totally coherent
$\lambda = 0$. Different shapes of the source were investigated in
the literature: $\Omega (Qr) = e^{-Qr}$; $e^{-Q^2r^2}$;
$1/(1+Qr)^2$; $[J_1(Qr)/(Qr)]^2$, to name a few \cite{Sources}.
The most advanced and complete discussion advocating this type of
$C_2(Q)$ and justifying its structure can be found in \cite{ALS}.

However, in a number of works it was strongly suggested that in
the case when a hadronizing source is {\it partially coherent} the
proper form of $C_2(Q)$ is the following one \cite{Weiner}:
\begin{equation}
C_2(Q) = 1 + 2p(1-p)\sqrt{\Omega(Q\cdot r)} + p^2 \Omega(Q\cdot
r)^2, \label{eq:second}
\end{equation}
with $r$ and $\Omega $ defined as above, where $p\in (0,1)$
replaces $\lambda$ (retaining, however, essentially its meaning).

The imminent question arises then: which expression is the proper
one? The answer was proposed in our paper \cite{KUW} were we have
shown that {\it both expressions are correct in its own way},
i.e., their form encodes information on the specific features of
hadronizing source which they describe; this point was not made
apparent in the previous works \cite{ALS,Weiner}. And so:
\begin{itemize}
\item Eq. (\ref{eq:first}) describes situation in which
hadronizing source can be regarded as consisting from the coherent
and chaotic subsources acting independently in the proportion
given by  the chaoticity parameter $\lambda$ \cite{ALS}.

\item Eq. (\ref{eq:second})  describes situation in which there is
only one hadronizing source but, for some reason, the phases of
all particles are partially aligned. This can happen, for example,
when hadronizing source is located in some constant external
field, as was the case considered in \cite{KUW}.
\end{itemize}
To summarize: the choice of one or other form of $C_2(Q)$
presented here amounts to making a nontrivial assumption
concerning the nature of the hadronizing source under
investigation. This should be at least remembered and
acknowledged, even if in practice only the first choice is
nowadays used \cite{CYW}.

\section{How to model BEC numerically}

The question of the numerical modelling of BEC is more important
than usually anticipated. The point is that to study multiparticle
production processes one uses numerical simulations, the Monte
Carlo event generators (MCEG) of different kinds. MCEG are based
on classical probabilistic schemes whereas BEC is, by definition,
quantum phenomenon and as such cannot be incorporated
straightforwardly into a MCEG. The suggested cure was the use of
the so called {\it afterburners}: one takes outcome of a given
MCEG and changes accordingly momenta of the selected identical
particles in such way as to fit the observed data on $C_2(Q)$.
However, it must be realized that by doing so one changes not only
the original energy-momenta and/or multiplicities (for which one
can correct later) but also (and usually unknowingly and in an
unknown way) the physics of the model used as the basis of the
MCEG chosen \cite{Foot1}.

The best solution would be to perform direct numerical simulations
in which MCEG would start from the input containing already
effects of BE statistics. What such input should be? The obvious
suggestion is: the one possessing property that particles
satisfying BE statistics tend to occupy in a maximal way the same
state, i.e., they exhibit a {\it bunching property}. This property
can be (at least in principle) modelled \cite{Idea}. There exist
already some examples of such effort. In \cite{ZajcMC} Metropolis
method was used with fully symmetrized wave function to convert
the set of $N$ uniformly distributed identical particles into the
set of $N$ particles exhibiting the effect of BEC. Closer
inspection shows that it happened because in this way particles
were effectively bunched in the phase space forming what
\cite{ZajcMC} called  {\it speckles}. In \cite{Cramer} one starts
with single particle and, using rejection method, adds to the
$N$-th particle the $(N+1)$-th one following the updated
probability as given by the fully symmetrized wave function for
$(N+1)$ particles. Again: the final distribution is characterized
by bunches of identical particles in the phase space. In
\cite{OMT} the main point was to account for the Negative Binomial
(NB) character of the observed multiplicity distributions $P(N)$
by assuming that particles of the same charge are most likely
being produced in the same cells into which the phase space has
been divided (it is rapidity in this work). All presented
algorithms are very time consuming (especially the first two) and
only the last one has been successfully applied to analysis of
$e^+e^-$ data \cite{OMT} (and never again).

In \cite{UWW} we have summarized our effort aimed at improving
this approach. From the examples mentioned it is clear that the
procedure of symmetrization of the initial set of identical
particles distributed somehow in the phase space takes too much
time to be of any practical use. On the other hand, it leads to
very interesting result, namely it shows that {\it this procedure
results in the effect of bunching of particles  in some regions of
phase space}. Because bunching is easier to simulate than
symmetrization, it is this phenomenon which we propose to use as
the cornerstone of the algorithm modelling BEC. Therefore we form
bunches (called by us {\it Elementary Emitting Cells} - EEC's) of
particles in energy. It can be shown that using Bose-Einstein (or
geometrical) form of distribution of particles in a single EEC one
gets the characteristic proper BE form of $<N(E)>$ together with
the characteristic shape of the $C_2(Q)$ function. When the
original energy distribution is thermal-like (exponential with
scale parameter $T$) then $T$ is temperature seen in $<N(E)>$,
whereas chemical potential present there is the main parameter
describing BE distribution of particles in EEC \cite{UWW}. The
picture proposed resembles closely a quantum version of the clan
model \cite{NB} (in which all particles in a clan are identical
bosons). We call it therefore {\it Quantum Clan Model} (QCM)
\cite{UWW}. In this case one gets final multiplicity distribution
in the form of P\'olya-Aeppli (geometric-Poisson) distribution
\cite{PA}, which differs from the NB distribution of \cite{NB}
only for very small multiplicities. The strength of BEC, as given
by parameter $\lambda$ in (\ref{eq:first}), is very sensitive to
the maximal allowed number of particles in the single EEC.
Therefore, for extremely high multiplicity cases, $\lambda$
exceeds $2$ (even for $C_2(Q)$), a fact not noticed before.

\section{Summary and conclusions}

Let us summarize points raised here.

\begin{itemize}
\item The first is that deciding, as it is usually done, on the
eq. (\ref{eq:first}) when addressing problem of BEC one {\it
tacitly assumes} that hadronizing source is not influenced by any
external field which could make it partially coherent (depending
on its strength). Therefore in situations where this cannot be
assured its better to use eq. (\ref{eq:second}). However, the
trouble is that so far its form is elaborated only for
two-particle BEC, multiparticle case still awaits its proper
treatment.

\item The second point is that using one of the {\it afterburners}
proposed in the literature in order to change the outcome of the
MCEG actually used, one accepts also (most times unknowingly!) all
changes in the physical picture underlying this MCEG. The only way
out would be to build a MCEG using the principle of BE statistics
as its corner stone. It can be done by endowing MCEG  from the
very beginning of the numerical simulation process with property
of bunching of identical particles (via geometrical, or
Bose-Einstein, particle distribution in each bunch assured
numerically). Effects of resonances, final state interaction and
the like, can be (in principle) accounted for. They always reduce
signal of BEC. The most difficult problem one encounters is the
implementation of corrections for nonconservation of
energy-momenta and charge introduced during the Monte Carlo
selection procedure used. We demonstrate that such program is
possible but, at the moment, still far from the completion.
\end{itemize}

\acknowledgements

GW would like to acknowledge support obtained from the
Bogolyubov-Infeld program in JINR and partial support of the
Ministry of Science and Higher Education under contracts
1P03B02230 and CERN/88/2006.


\begin{thebibliography}{99}
\bibitem{Foot1} In \cite{UWW} it was demonstrated using simple version of cascade model
                that forcing BEC in this way results in the multicharged vertices to be
                formed in the cascade process whereas the original cascade had only single
                valued vertices. Effects of this kind should be therefore expected when using
                MCEG based on different physical principles. To the best of our knowledge
                this problem has never been investigated.
\end{thebibliography}
\end{document}